\documentclass[12pt]{article}
\usepackage{amsmath}
\usepackage[round]{natbib}
\usepackage[english]{babel}
\usepackage[utf8]{inputenc}
\usepackage{accents}
\usepackage{graphicx}
\usepackage{rotating}
\usepackage{pdflscape}
\usepackage{float}

\usepackage{epsfig}
\usepackage{setspace} 
\usepackage{array}
\usepackage{multirow}
\usepackage{color}
\newcolumntype{L}[1]{>{\raggedright\let\newline\\\arraybackslash\hspace{0pt}}m{#1}}
\newcolumntype{C}[1]{>{\centering\let\newline\\\arraybackslash\hspace{0pt}}m{#1}}
\newcolumntype{R}[1]{>{\raggedleft\let\newline\\\arraybackslash\hspace{0pt}}m{#1}}

\begin{document}

\title{A comparison of the discrimination performance of lasso and maximum likelihood estimation in logistic regression models}

\author{Gilberto P. Alcântara Junior$^1$ \and Gustavo H. A. Pereira$^1$}
\date{}

\maketitle

\vspace{-11.5mm} \noindent 
\begin{center}
$^1$Department of Statistics, Federal University of São Carlos, São Carlos, Brazil \\
\end{center}
\vspace{1mm}

\begin{abstract}
Logistic regression is widely used in many areas of knowledge. Several works compare the performance of lasso and maximum likelihood estimation in logistic regression. However, part of these works do not perform simulation studies and the remaining ones do not consider scenarios in which the ratio of the number of covariates to sample size is high. In this work, we compare the discrimination performance of lasso and maximum likelihood estimation in logistic regression using simulation studies and applications. Variable selection is done both by lasso and by stepwise when maximum likelihood estimation is used.  We consider a wide range of values for the ratio of the number of covariates to sample size. The main conclusion of the work is that lasso has a better discrimination performance than maximum likelihood estimation when the ratio of the number of covariates to sample size is high.

\end{abstract}

\vspace{0mm} \noindent {\textbf{Keywords}}: lasso, logistic regression, maximum likelihood estimation, stepwise.

\vspace{3mm}

\section{Introduction}
\label{sec:introd}

Logistic regression is a widely used model for binary responses. 
\hspace{-0.45cm} Part of its success lies in the fact that it has interpretable parameters. The parameters in logistic regression are usually estimated by the maximum likelihood method \citep{hosmer2013applied}. Stepwise is commonly used for variable selection \citep{gaebel2016predictors,muller2020development}.

Lasso \citep{tibshirani1996regression} is an estimation method that can be used in many regression models. It is often used when prediction is the main purpose of model development, because it usually produces better predictions than traditional methods \citep{hastie2019statistical}. In addition, unlike the maximum likelihood method, it can be used when the number of covariates is greater than the number of observations. Another interesting feature of lasso is that it also performs variable selection, because the estimates of several parameters are usually zero.

Several works compare lasso, the maximum likelihood method and other methods using real datasets with binary response (see for example \citet{steyerberg2000prognostic}, \citet{greenwood2020comparison} and \citet{liu2021machine}). In general, they concluded that lasso has a better predictive performance than maximum likelihood estimation, especially when the sample size is small compared to the number of covariates.

Few works compare lasso and the maximum likelihood method using extensive simulation studies in logistic regression.  \citet{van2020regression}, \citet{riley2021penalization} and \citet{martin2021developing} compared these methods and others. They concluded that lasso and other shrinkage methods do not guarantee improved performance. However, they did not consider scenarios in which the ratio of the number of covariates to sample size is greater than 0.2. In addition, they focused on the estimation of the probability of the response assuming each outcome instead of studying the discrimination performance of the methods. In many real problems, the goal is the latter and not the former. Finally, these works did not evaluate the maximum likelihood method associated with a variable selection method as it is usually used in practice. \citet{pavlou2016review} and \citet{leeuwenberg2022performance} also compared lasso, the maximum likelihood estimation and other methods in logistic regression using simulation studies. However, the authors considered only settings with low dimension.

In this work, we perform extensive simulation studies to compare the discrimination performance of lasso and the combination of stepwise with maximum likelihood estimation in logistic regression models. We also include in the analyses the combination of lasso and maximum likelihood method, in which the latter is used for parameter estimation and the former for variable selection. In the simulation studies, we vary the ratio of number of covariates to sample size from 0.01 to 0.5.
The performance of these methods is also compared using nine real datasets.

The remainder of this paper is organized as follows. Section 2 presents the model and the methods considered in this work. The discrimination performance of the methods are compared using Monte Carlo simulation studies and nine real databases in Section 3 and 4, respectively. Concluding remarks are provided in Section 5. 


\section{Methodology}
\label{sec:meth}

Let $y_1, y_2, \ldots, y_n$ be independent random variables, where each $y_i$, $i=1,2,\ldots,n$, is Bernoulli distributed with probability of success $\mu_i$. The logistic regression model for binary responses is defined as 
\begin{equation}
\label{eq:logist}
g(\mu_i)=\log\left(\frac{\mu_i}{1-\mu_i}\right)= \alpha + x_{i}^\top\beta,
\end{equation}
where $x_{i}=(x_{i1},x_{i2},\ldots,x_{ip})^\top$ is a vector of known covariates, $\alpha$ is an unknown intercept  parameter and $\beta=(\beta_1,\beta_2,\ldots,\beta_p)^\top$ is a vector of unknown parameters ($\beta \in R^p$). Logistic regression is a particular case of generalized linear models in which the response variable is Bernoulli distributed and $g(.)$ is the logit link function \citep{mcc+nel89}. 

The parameters of the logistic regression model can be estimated by maximum likelihood, in which estimators of the parameters are obtained maximizing the log-likelihood of the model (\ref{eq:logist}). This maximization uses a  
nonlinear optimization algorithm, such as the iterative weighted least squares \citep[Section 2.5]{mcc+nel89}. The log-likelihood of the model (\ref{eq:logist}) is given by
\begin{equation}
\label{eq:loglike}
l(\alpha,\beta) = \sum_{i=1}^n[y_i\log(\mu_i) + (1-y_i)\log(1-\mu_i)]
\end{equation}
where $\mu_i = e^{\alpha + x_{i}^\top\beta}/(1 + e^{\alpha + x_{i}^\top\beta})$. In the first method considered here, stepwise is used for variable selection and maximum likelihood is used for parameter estimation. We denote it by StepML.

In the lasso method, the parameters are estimated by minimizing the following function:
\begin{equation}
\label{eq:lasso}
-l(\alpha,\beta) + \lambda\sum_{j=1}^p|\beta_j|,
\end{equation}
where $\lambda$ is a tuning parameter. In this work, we chose the tuning parameter using ten-fold cross validation \citep[page 614]{fischetti2017r}. The chosen $\lambda$ was the one that maximized the average of the area under the ROC curve (AUC) in the validation datasets.

As lasso also selects covariates, it is reasonable to use it for variable selection and another method for parameter estimation. However, to the best of our knowledge, there is no work that compares the combination of lasso and a parameter estimation method with other techniques in logistic regression. Here, we consider the combination of lasso for variable selection and maximum likelihood for parameter estimation. We denote this combination as LassoML.


\section{Simulation studies}
\label{sec:simul}

To compare the discrimination performance of the methods described in Section \ref{sec:meth} in the logistic regression, we performed a Monte Carlo simulation study. Following \citet{van2020regression}, we considered a full factorial simulation setup varying the following factors: number of covariates ($p=10, 30 \text{ and } 50$), outcome rate event or percentage of successes (20\% and 50\%) and the correlation between predictors. The covariates were generated as random draws from the multivariate normal distribution with mean vector composed of zeros, variance vector composed of ones and correlations given by $(-\rho)^{|i-j|}$ \citep{hastie2020best}, where $i$ and $j$ are the $ith$ and $jth$ covariates, respectively, and $\rho = 0.5$ in the half of the scenarios and  $\rho = 0.9$ in the remaining ones.

For each scenario, we considered that the $80\%$ first covariates were noise ones (associated parameters equal to zero) and the value of the parameter for the $(0.8p+i)th$ covariate $(i=1,2,\hdots,p-0.8p)$ was given by $(-1)^{(i+1)}(0.5)$. The following sample sizes were considered: $n = 100, 200, 500 \text{ and } 1000$. Each sample was split in training dataset ($70\%$) and test dataset ($30\%$) and 500 Monte Carlo replications were considered in each scenario and sample size.

We used the Gini coefficient (GC) \citep{thomas2017credit} as the measure of the discrimination performance. This measure is a transformation of the area under the ROC curve given by 
$\text{GC } = 2(\text{AUC } - 0.5)$. The GC is usually used to evaluate credit scoring models \citep{silva2022class} instead of AUC, because it takes values in the interval $(0,1)$. 

\begin{figure}[ht!]
\centering 
\includegraphics[width=16cm]{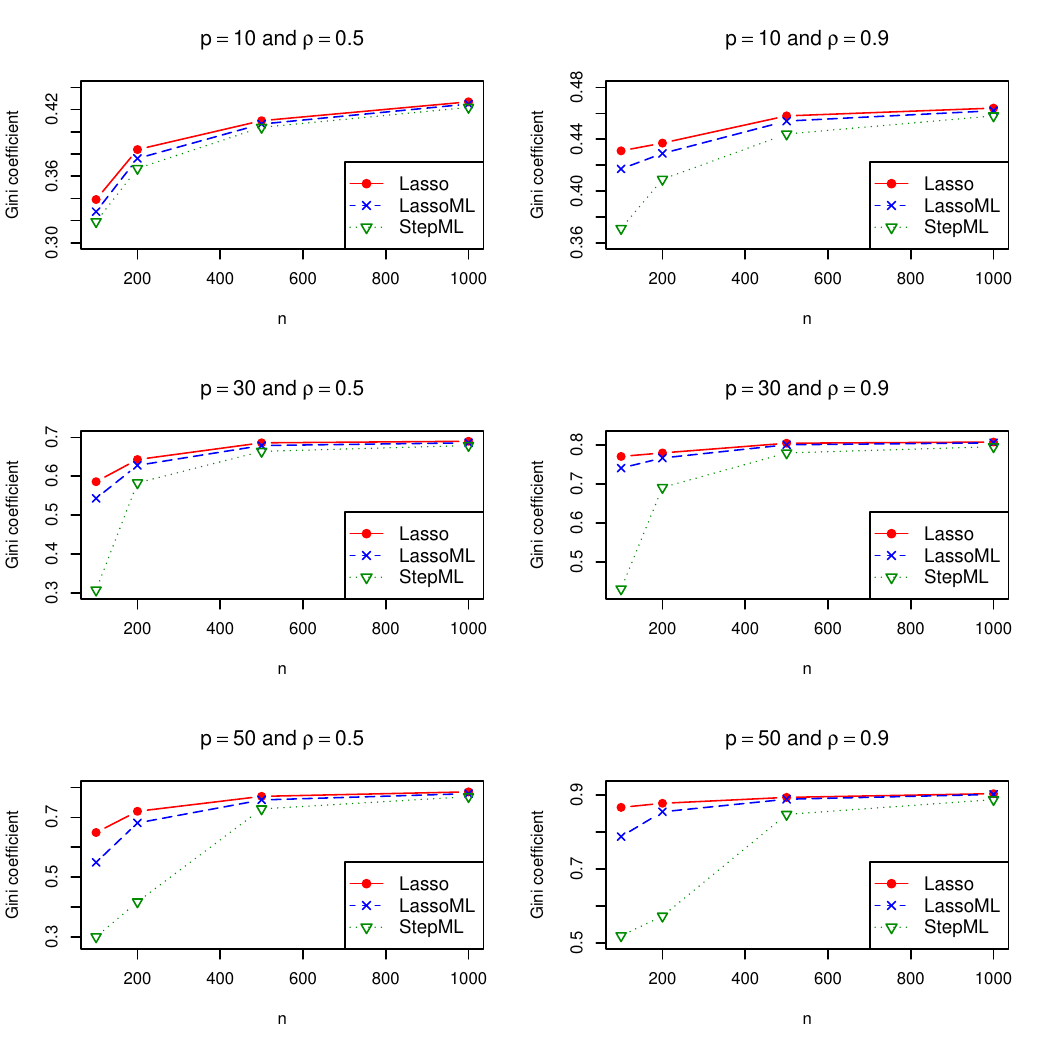} 
\caption{Average Gini coefficient for scenarios with outcome rate event equals to $50\%$}
\label{fig:sim1}
\end{figure}

Figure \ref{fig:sim1} presents the results of the simulations for the scenarios in which the outcome rate event (ORE) is equal to 0.5. When $p=50$ and $n=100$, the average GC is much higher for Lasso than for StepML (0.65 versus 0.30 when $\rho = 0.5$). The average GC for LassoML is between the other two methods but closer to Lasso (0.55 when $\rho = 0.5$). The difference in the discrimination performance of the methods is related especially to the value of the ratio $p/n$. In general, the greater this ratio, the greater the difference in the average GC between the three methods and Lasso is the method of best discrimination performance followed by LassoML. When $p/n$ is less than 0.05, the three methods present similar performance. 
\hspace{-0.35cm} The difference in the average GC between the methods is slightly
higher when $\rho$ is changed from $0.5$ to $0.9$. As the differences are small when this change is made, the level of correlations between the covariates does not considerably affect the relative discrimination performance of the methods considered here.

The results of the simulations for the scenarios in which the ORE is equal to 0.2 are shown in Figure 2. When we change the ORE from $0.5$ to $0.2$, the conclusions are similar in most of the cases. However, for fixed $p$, $n$ and $\rho$, in some case, the difference between the methods is considerably greater for 
ORE equals to 0.2 than when it is 0.5. For example, When $p=50$, $n=500$ and $\rho=0.9$, the three methods have similar discrimination performance when the ORE is equal to 0.5. However, for the same $p$, $n$ and $\rho$ and ORE equals 0.2, the average GC is considerably greater for Lasso (0.91) than for StepML (0.75).

\begin{figure}[ht!]
\centering 
\includegraphics[width=16cm]{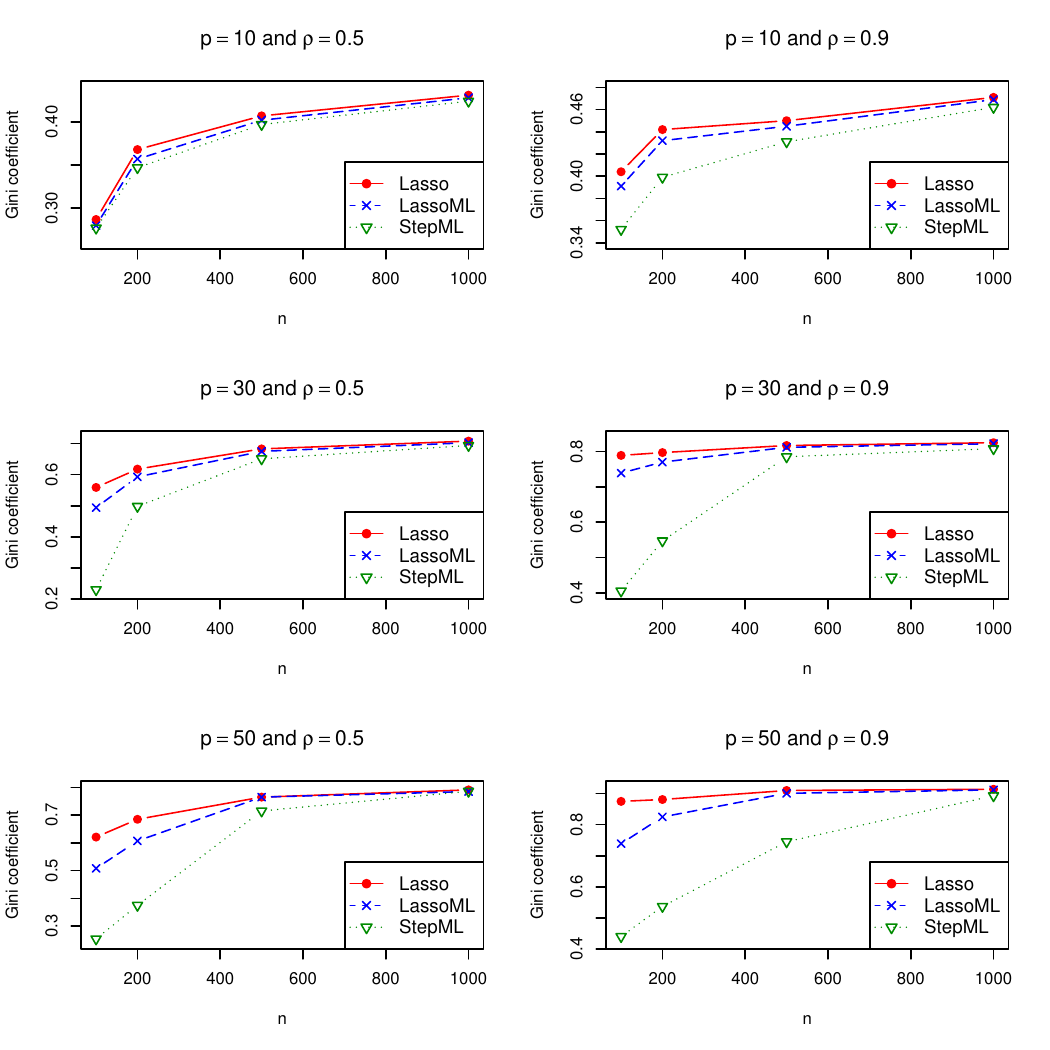} 
\caption{Average Gini coefficient for scenarios with outcome rate event equals to $20\%$}
\label{fig:sim2}
\end{figure}

\section{Applications}
\label{sec:app}

We also compared the methods considered here using nine applications. The datasets were obtained from different areas and have different values of $n$, $p/n$ and ORE. Table \ref{ta:datasets} presents the characteristics  of these datasets. Note that $p/n$ varies from 0.001 to 4.783.

 \begin{table} [ht!]   
 \caption {Features of the used datasets.} 
     \vspace{-0.1cm} 
    \begin{center}
\begin{tabular} {crrccc}  
\hline
Dataset	&	$n$	&	$p$	&	$p/n$	&	ORE $(\%)$	&	Reference	\\
\hline
1	&	30000	&	23	&	0.001	&	22	&	\citet{yeh2009comparisons}	\\
2	&	3656	&	15	&	0.004	&	15	&	\citet{detrano1989international}	\\
3	&	392	&	8	&	0.020	&	33	&	\citet{ramana2011critical}	\\
4	&	123	&	6	&	0.049	&	50	&	\citet{thrun1991monk}	\\
5	&	569	&	30	&	0.053	&	37	&	\citet{street1993nuclear}	\\
6	&	351	&	34	&	0.097	&	36	&	\citet{sigillito1989classification}	\\
7	&	195	&	22	&	0.113	&	75	&	\citet{little2007exploiting}	\\
8	&	70	&	205	&	2.929	&	41	&	\citet{zarchi2018scadi}	\\
9	&	115	&	550	&	4.783	&	33	&	\citet{sorlie2003repeated}	\\
\hline
    \end{tabular}
\end{center}
    \label{ta:datasets}
\end{table}

For each application, we split the dataset in training dataset $(70\%)$ and test dataset $(30\%)$ and repeated this procedure 100 times. Table \ref{ta:dataresults} presents the average and standard deviation of the GC for the three methods in the test datasets.

 \begin{table} [ht!]   
 \caption {Average and standard deviation of the Gini coefficient for the nine applications.} 
     \vspace{-0.1cm} 
    \begin{center}
\begin{tabular} {cccc}  
\hline
Dataset	&	Method	&	\multicolumn{2}{c}{Gini coefficient}			\\
\cline{3-4}
	&		&	Average	&	Standard deviation	\\
\hline							
	&	Lasso	&	0.445	&	0.012	\\
1	&	LassoMV	&	0.446	&	0.012	\\
	&	StepMV	&	0.445	&	0.012	\\
\hline							
	&	Lasso	&	0.462	&	0.033	\\
2	&	LassoMV	&	0.461	&	0.033	\\
	&	StepMV	&	0.460	&	0.033	\\
\hline							
	&	Lasso	&	0.693	&	0.061	\\
3	&	LassoMV	&	0.694	&	0.061	\\
	&	StepMV	&	0.690	&	0.061	\\
\hline							
	&	Lasso	&	0.528	&	0.133	\\
4	&	LassoMV	&	0.530	&	0.181	\\
	&	StepMV	&	0.529	&	0.132	\\
\hline							
	&	Lasso	&	0.984	&	0.013	\\
5	&	LassoMV	&	0.967	&	0.034	\\
	&	StepMV	&	0.892	&	0.038	\\
\hline							
	&	Lasso	&	0.813	&	0.065	\\
6	&	LassoMV	&	0.784	&	0.075	\\
	&	StepMV	&	0.710	&	0.095	\\
\hline							
	&	Lasso	&	0.784	&	0.079	\\
7	&	LassoMV	&	0.776	&	0.089	\\
	&	StepMV	&	0.781	&	0.102	\\
\hline							
	&	Lasso	&	0.662	&	0.183	\\
8	&	LassoMV	&	0.591	&	0.197	\\
	&	StepMV	&	0.243	&	0.239	\\
\hline							
	&	Lasso	&	0.444	&	0.120	\\
9	&	LassoMV	&	0.335	&	0.164	\\
	&	StepMV	&	0.115	&	0.145	\\

\hline
    \end{tabular}
\end{center}
    \label{ta:dataresults}
\end{table}

The conclusions are similar to those obtained in the simulation studies. As observed in that analysis, the difference in the discrimination performance between the methods is related especially to the value of the ratio $p/n$. For the the two datasets in which $p > n$, the average GC is much higher for Lasso than for  StepML and LassoML has also an average GC much higher than StepML and lower than Lasso. On the other hand, in the four datasets in which $p/n$  is lower than 0.05, the discrimination performance of the three methods is similar. The other three datasets have $p/n$ between $0.05$ and $0.12$. In two of them, the average GC follows the same order across the methods noted in the datasets in which $p > n$. However, the differences in the average GC are lower for these datasets because $p/n$ is not so high. The last dataset (number 7) has an unexpected results. It presents $p/n$ greater than 0.1, but the average GC is similar for the three methods. For the datasets in which Lasso has a higher average GC than the other methods, it also presents a lower standard deviation of the GC.

\section{Concluding remarks}
\label{sec:conc}

In this work, we compared the discrimination performance of three methods used to fit a logistic regression model. The considered methods are the following: lasso, the combination of stepwise for variable selection and maximum likelihood for parameter estimation and the combination of lasso for variable selection and maximum likelihood for parameter estimation.  These methods were compared using Monte Carlo simulation studies and nine applications. The main conclusion of the work is that lasso has 
a better discrimination performance than the other two methods when the ratio of the number of covariates $(p)$ to sample size $(n)$ is high. 

The analyses performed here suggest that the three methods have similar discrimination performance when $p/n$ is lower than $0.05$. For values higher than $0.05$, in general, the greater $p/n$, the greater the difference in the discrimination performance between lasso and the other two methods. 
The relative performance of the methods seems to be less affected by the outcome rate event and by the level of correlation between the covariates. However, in general, the superiority of lasso compared to the other methods seems to be slightly higher when the outcome rate event is farther from 0.5 and when the correlation between the covariates is higher.

Considering all the analyses performed in this work, lasso did not present a lower discrimination performance than the other methods in any application or scenario of the simulation studies. In addition, lasso is much better than the other methods considered here when $p/n$ is high. Therefore, if the main goal of a work is obtaining a model with good discrimination performance, the logistic regression model should be fitted using lasso instead of maximum likelihood estimation.

\section{Acknowledgments}

This study was financed in part by the Coordenação de Aperfeiçoamento de Pessoal de Nível Superior - Brasil (CAPES) - Finance Code 001.


\singlespacing   

\bibliographystyle{elsarticle-harv}


\end{document}